\documentclass[12pt]{iopart}
\usepackage{graphicx} 

\begin{document}

\title{How the Scalar Field of Unified Dark Matter Models Can Cluster}

\author{Daniele Bertacca}
\address{Dipartimento di Fisica Teorica and Dipartimento di Fisica Generale
``Amedeo Avogadro'', Universit\`a di Torino, 
and INFN Sezione di Torino, via P. Giuria 1, I-10125 Torino, Italy}
\ead{bertacca@to.infn.it}

\author{Nicola Bartolo}
\address{Dipartimento di Fisica ``Galileo Galilei", Universit\`a di Padova, 
and INFN Sezione di Padova, via F. Marzolo, 8 I-35131 Padova Italy}
\ead{nicola.bartolo@pd.infn.it}

\author{Antonaldo Diaferio}
\address{ Dipartimento di Fisica Generale ``Amedeo Avogadro'',
Universit\`a di Torino and INFN Sezione di Torino, via P. Giuria 1,
 I-10125 Torino, Italy}
\ead{diaferio@to.infn.it}

\author{Sabino Matarrese}
\address{Dipartimento di Fisica ``Galileo Galilei", Universit\`a di Padova, 
and INFN Sezione di Padova, via F. Marzolo, 8 I-35131 Padova Italy}
\ead{sabino.matarrese@pd.infn.it}

\begin{abstract}
We use scalar-field Lagrangians with a non-canonical
kinetic term to obtain unified dark matter models where both 
the dark matter and the dark energy, the latter mimicking a cosmological 
constant, are described by the scalar field itself. In this framework, 
we propose a technique to reconstruct models where the effective 
speed of sound is small enough that the scalar field can cluster. 
These models avoid the strong time evolution 
of the gravitational potential and the large Integrated Sachs-Wolfe 
effect which have been a serious drawback of previously considered models. 
Moreover, these unified dark matter scalar field models can be easily 
generalized to behave as dark matter plus a dark energy component  
behaving like any type of quintessence fluid.
\end{abstract}
\maketitle

\section{Introduction}

If we assume that General Relativity correctly describes the phenomenology
of our universe, astronomical observations provide compelling evidence 
that (1) the dynamics of cosmic structures is dominated by dark matter (DM), 
a cold collisionless component mostly made of hypothetical elementary 
particles, and (2) the expansion of the universe is currently 
accelerating because of the presence of a positive cosmological constant 
or a more general Dark Energy (DE) component.
The DM particles have not yet been detected and there is no theoretical 
justification for the tiny cosmological constant (or more general DE 
component) implied by observations (see, e.g. 
Refs.~\cite{Spergel:2006hy,Dunkley:2008ie}). 
Therefore, over the last decade, the search 
for extended or alternative theories of gravity has flourished.

In this paper we focus on unified models of DM and DE
(UDM), in which a single scalar field provides an alternative 
interpretation to the nature of the dark components of the universe.
Compared with the standard DM + DE models (e.g.
$\Lambda$CDM), these models have the advantage 
that one can describe the dynamics of the universe
with a single dark fluid which triggers both the accelerated expansion 
at late times and the large-scale structure formation at earlier times.
Moreover, for these models, we can use
Lagrangians with a non-canonical kinetic term, namely a
term which is an arbitrary function of the square of the time derivative 
of the scalar field, in the homogeneous and isotropic background. 
These models are known as ``k-essence models''
\cite{Chiba:1999ka,AMS1,AMS2,Rendall:2005fv,Fang:2006yh,Babichev:2007dw}
(see also \cite{Li:2006bx,Kang:2007vs,Babichev:2006cy,Babichev:2007tn,Bazeia:2007df,Calcagni:2006ge}) 
and have been inspired by earlier studies of k-inflation 
\cite{kinf} \cite{Garriga:1999vw} 
(a complete list of dark energy models can be found in the review 
\cite{Copeland:2006wr}). 

Most UDM models studied so far in the literature 
require non-trivial fine tunings.  
%and do not solve the 
%so-called cosmic coincidence problem, namely why $\Omega_{\rm DM}$   
%and $\Omega_{\Lambda}$ are both of order unity today. 
Moreover, the viability of UDM models strongly depends on the 
value of the effective speed of sound $c_s$ 
\cite{Huss,Garriga:1999vw,Mukhanov:2005sc}, which has to be small enough 
to allow structure formation  
\cite{Sandvik:2002jz,Giannakis-Hu} and   
to reproduce the observed pattern of Cosmic Microwave Background (CMB) 
temperature anisotropies \cite{Huss,Bertacca:2007ux,Sandvik:2002jz,Finelli,Bertacca:2007cv}. 
The prospects for a unified description of DM/DE (and inflation) through a 
single scalar field has been addressed also in Ref.~\cite{LiddleLopez}. \\

Several adiabatic or, equivalently, 
purely kinetic models have been investigated in the literature:   
for example, the generalized 
Chaplygin gas \cite{Kamenshchik:2001cp,Bilic02,Bento:2002ps} (see also
Refs.~\cite{GCG,chap_con1,chap_con2,chap_con3,chap_con4,Giannantonio:2006ij}),
the Modified Chaplygin gas \cite{MCG}, the Scherrer \cite{Scherrer:2004au} 
and generalized Scherrer solutions \cite{Bertacca:2007ux}, 
the single dark perfect fluid with a simple 2-parameter barotropic 
equation of state \cite{Bruni}, or the homogeneous scalar field deduced 
from the galactic halo space-time \cite{DiezTejedor:2006qh} (see also
Ref.\cite{Bertacca:2007fc}). \\ 

Moreover, one can build up scalar field models where the 
constraint that the Lagrangian is constant along the classical trajectories,
namely the solutions of the equations of motion, allows to describe a 
UDM fluid whose average behaviour is that of dark matter plus a cosmological 
constant \cite{Bertacca:2007ux} 
(see also Ref.~\cite{stability2,stability1,prc}, 
for a different approach). 
Alternative approaches to the unification of DM 
and DE have been proposed in Ref.~\cite{takayana}, 
in the framework of supersymmetry, in Ref.\cite{LukesGerakopoulos:2008rr} 
in connection with chaotic scalar field solutions in Friedmann-Robertson-Walker
cosmologies and in Ref.~\cite{bono}, 
in connection with the solution to the strong CP problem.
One could also easily reinterpret UDM models based on a scalar field 
Lagrangian in terms of generally non-adiabatic fluids 
\cite{DiezTejedor:2005fz,Brown:1992kc} (see also \cite{Bilic:2008zk}). 

Here we choose to investigate the class of scalar-field Lagrangians 
with a non-canonical kinetic term to obtain UDM models.
In Ref.~\cite{Bertacca:2007ux}, the authors require that the 
Lagrangian of the scalar field is constant along the classical trajectories.
Specifically, by requiring that $\mathcal{L}=-\Lambda$ on cosmological scales, 
the background they obtain is identical to the background of the 
$\Lambda$CDM model.
In this case the limited number of degrees of freedom does not leave 
any room for choosing the evolution of the effective speed of sound 
$c_s^2$ in agreement with observations \cite{Bertacca:2007ux}. 

Moreover, one of the main issues of these UDM models is to see whether their 
single dark fluid is able to cluster 
and produce the cosmic structures we observe in the universe today. 
In fact, the effective speed of sound can 
be significantly different from zero at late times; the corresponding 
Jeans length (or sound horizon), below which the dark fluid can not cluster,
can be so large that the gravitational potential first strongly 
oscillates and then decays \cite{Bertacca:2007cv}, thus preventing structure 
formation. 
Previous work attempted to solve this problem by a severe
fine-tuning of the parameters appearing in the Lagrangian (see for example
\cite{chap_con1,chap_con2,chap_con3,chap_con4,Scherrer:2004au,Giannakis-Hu}).

In Section \ref{basics}, we layout the basic equations; 
in Sections \ref{UDM}, \ref{BIL} and \ref{UDMg(X)-generic}, we suggest a reconstruction technique 
to find models where the effective speed of sound is small enough 
that the scalar field can cluster. Specifically, in Section \ref{BIL} 
we consider a model with kinetic term of Born-Infeld type 
\cite{prc,abramo1,abramo2,Alishahiha:2004eh,stability1,stability2}
that does not allow a strong time evolution 
of the gravitational potential and the large Integrated Sachs-Wolfe (ISW) 
effect which have been a serious drawback of previous models.
In Section \ref{Class-eq}, we consider a more general class of UDM 
Lagrangians, with a non-canonical kinetic term, whose 
equations of motion are dynamically equivalent to those of the 
previous models. 
Finally, in Section \ref{GUDM} we study a possible way to generalize 
UDM models so that they can mimic dark matter and dark energy 
in the form of a general quintessence fluid. 

\section{Basic equations}
\label{basics}
The action describing the dark matter unified models can be written as
\begin{equation}\label{eq:action}
S = S_{G} + S_{\varphi}= 
\int d^4 x \sqrt{-g} \left[\frac{R}{2}+\mathcal{L}(\varphi,X)\right] \, ,
\end{equation}
where 
\begin{equation}\label{x}
X=-\frac{1}{2}\nabla_\mu \varphi \nabla^\mu \varphi \;.
\end{equation}
We use units such that $8\pi G = c^2 = 1$ and signature $(-,+,+,+)$
(greek indices run over spacetime dimensions, whereas latin indices label 
spatial coordinates).\\
The energy-momentum tensor of the scalar field $\varphi$ is
\begin{equation}
  \label{energy-momentum-tensor}
  T^{\varphi}_{\mu \nu } = 
- \frac{2}{\sqrt{-g}}\frac{\delta S_{\varphi }}{\delta
    g^{\mu \nu }}=\frac{\partial \mathcal{L}
(\varphi ,X)}{\partial X}\nabla _{\mu }\varphi
  \nabla _{\nu }\varphi +\mathcal{L}(\varphi ,X)g_{\mu \nu }.
\end{equation}
If $X$ is time-like, $S_{\varphi}$ describes a perfect fluid with 
$T^{\varphi}_{\mu \nu }=(\rho +p) u_{\mu} 
u_{\nu }+p\,g_{\mu \nu }$, with pressure 
\begin{equation}
  \label{pressure}
  \mathcal{L}=p(\varphi ,X)\;,  
\end{equation}
and energy density
\begin{equation}
  \label{energy-density}
  \rho =\rho 
(\varphi ,X)= 2X\frac{\partial p(\varphi ,X)}
  {\partial X}-p(\varphi ,X)  
\end{equation}
where 
\begin{equation}
  \label{eq:four-velocity}
  u_{\mu }= \frac{\nabla _{\mu }\varphi }{\sqrt{2X}}
\end{equation}
is  the four-velocity.

Now we assume a flat, homogeneous Friedmann-Robertson-Walker 
background with scale factor $a(t)$. 
With this metric, when the energy density of the radiation
becomes negligible, and disregarding also the small baryonic component, 
the background evolution of the universe is 
completely described by the following equations 
\begin{equation}
\label{eq_u1}
H^2 =\frac{1}{3}  \rho\, ,
\end{equation}
\begin{equation}
\label{eq_u2}
\dot{H} = - \frac{1}{2} (p + \rho)\, ,
\end{equation}
where the dot denotes differentiation w.r.t. the cosmic time $t$ and 
$H=\dot{a}/a$. In these equations, the energy density and pressure 
of our scalar field $\varphi$, are supposed to describe both the dark matter 
and dark energy fluids. 

On the background, the kinetic term becomes $X=\frac{1}{2}\dot{\varphi}^2$, 
and the equation of motion for the homogeneous mode $\varphi(t)$ reads
\begin{equation}
 \label{eq_phi}
\left(\frac{\partial p}{\partial X} 
+2X\frac{\partial^2 p}{\partial X^2}\right)\ddot\varphi
+\frac{\partial p}{\partial X}(3H\dot\varphi)+
\frac{\partial^2 p}{\partial \varphi \partial X}\dot\varphi^2
-\frac{\partial p}{\partial \varphi}=0 \;. 
\end{equation}
The two relevant relations for the dark energy problem are the 
equation of state $w \equiv p/\rho$, which, in our case, reads 
\begin{equation}
\label{w}
w = \frac{p}{2X \frac{\partial p}{\partial X} - 
p} \;, 
\end{equation} 
and the effective speed of sound 
\begin{equation}
\label{cs}
c_s^2 \equiv  \frac{(\partial p /\partial X)}
{(\partial \rho/\partial X)} = 
\frac{\frac{\partial p }{\partial X}}{\frac{\partial p}
{\partial X}+ 2X\frac{\partial^2 p}{\partial X^2}} \;. 
\end{equation}
The latter relation plays a major role in the evolution of the scalar field
perturbations $\delta\varphi$ and in the growth of the 
overdensities $\delta\rho$.
In fact, we start from small inhomogeneities of the scalar field 
$\varphi(t,x)=\varphi_0(t)+\delta\varphi(t,\mbox{\boldmath $x$})$, and
write the metric in the longitudinal gauge, 
\begin{equation}
ds^2 = - (1+2 \Phi)dt^2 + a^2(t)(1-2 \Phi) \delta_{ij} dx^i dx^j \;,
\end{equation}
(where $\delta_{ij}$ is the Kronecker symbol), having  
used the fact that $\delta T_{i}^j = 0$ 
for $i \neq j$~\cite{Mukhanov:1990me}; here $\Phi$ is
the peculiar gravitational potential. When
we linearize the $(0-0)$ and $(0-i)$ components of Einstein equations 
(see Ref.~\cite{Garriga:1999vw} and Ref.~\cite{Mukhanov:2005sc}), we obtain
the second order differential equation 
\cite{Mukhanov:2005sc} \cite{Bertacca:2007cv}
\begin{equation}
\label{diff-eq_u}
u''-c_s^2 \nabla^2 u - \frac{\theta''}{\theta}u=0\;
\end{equation}
where primes indicate derivatives w.r.t. the conformal time $\eta$, defined 
through $d\eta = dt/a$; $u\equiv 2 \Phi/(p+\rho)^{1/2}$ and  
$\theta \equiv (1+p/\rho)^{-1/2}/(\sqrt{3}a)$ \cite{Mukhanov:2005sc}.

One of the main issues in the framework of UDM model building 
is to see whether the single dark fluid is able to cluster 
and produce the cosmic structures we observe in the universe. 
In fact, the sound speed appearing in Eq.~(\ref{diff-eq_u}) can
be significantly different from zero at late times; the corresponding 
Jeans length (or sound horizon), below which the dark fluid can not cluster,
can be so large that the gravitational potential first strongly 
oscillates and then decays \cite{Bertacca:2007cv}, thus 
preventing structure formation. 

Previous work attempted to solve this problem by a severe
fine-tuning of the parameters appearing in the Lagrangian (see for example
\cite{chap_con1,chap_con2,chap_con3,chap_con4,Scherrer:2004au,Giannakis-Hu,Bertacca:2007ux,Bertacca:2007cv}).
Here, we propose a class of UDM models where, 
at all cosmic times, the sound speed is 
small enough that cosmic structure can form.
To do so, a possible approach is to consider a scalar field
Lagrangian $\mathcal{L}$ of the form 
\begin{equation}
 \label{Lagrangian}
 \mathcal{L}= p(\varphi ,X)=f(\varphi)g(X)-V(\varphi) \; .  
\end{equation}
In particular, by introducing the two potentials $f(\varphi)$ and $V(\varphi)$,
we want to decouple the equation of state parameter $w$ and the sound speed 
$c_s$. This condition does not occur 
when we consider either Lagrangians with purely kinetic terms or 
Lagrangians like $\mathcal{L}=g(X)-V(\varphi)$ or 
$\mathcal{L}=f(\varphi)g(X)$~(see for example \cite{Bertacca:2007ux}).

Actually, we could start from a more general Lagrangian where 
$g=g(h(\varphi)X)$. However, by defining a new kinetic
term $Y=h(\varphi)X$, $h(\varphi)$ disappears and 
we need to recast $w$  and $c_s$  in terms of the new kinetic term. 
Therefore, this generalization does not describe 
a kinematics different from that generated by Eq. (\ref{Lagrangian}) 
(see Section \ref{Class-eq}, \ref{GeneralClass-eq} and \ref{CoordTransform}).

In the following sections we will describe how 
to construct UDM models based on Eq.~(\ref{Lagrangian}).

\section{How to construct UDM models}
\label{UDM}
Let us consider the scalar field Lagrangian of Eq. (\ref{Lagrangian}).
The energy density $\rho$, the equation of state $w$ and 
the speed of sound $c_s^2$ are
\begin{equation}
\label{rho_fg-v}
\rho(X,\varphi)=f(\varphi)\left[2X\frac{\partial g(X)}{\partial X}
-g(X)\right]-V(\varphi)\;,
\end{equation}
\begin{equation}
\label{w_fg-v}
w(X,\varphi)= \frac{f(\varphi)g(X)-V(\varphi)}{f(\varphi)
\left[2X \left(\partial g(X)/\partial X\right) -g(X)\right] -V(\varphi)}\;,
\end{equation}
\begin{equation}
\label{c_s^2_fg-v}
c_s^2(X)=\frac{\left(\partial g(X)/\partial X\right)}{\left(\partial g(X)/
\partial X\right)+ 2X\left(\partial^2 g(X)/\partial X^2\right)} \;, 
\end{equation} 
respectively.  The equation of motion (\ref{eq_phi}) becomes
\begin{equation}
 \label{eq_phi_fg-v}
\fl
\left(\frac{\partial g}{\partial X} 
+2X\frac{\partial^2 g}{\partial X^2}\right)\frac{d X}{d N}
+6X\frac{\partial g}{\partial X}+
\frac{d \ln f}{d N}\left(2X\frac{\partial g}{\partial X}
- g\right)
-\frac{1}{f}\frac{d V}{d N}=0\;,
\end{equation}
where $N=\ln a$.

Unlike models with a Lagrangian with
purely kinetic terms, here we have one more degree of freedom, 
the scalar field configuration itself.
Therefore this allows to impose a new condition 
to the solutions of the equation of motion.
In Ref.~\cite{Bertacca:2007ux}, the scalar field Lagrangian was required 
to be constant along the classical trajectories.
Specifically, by requiring that $\mathcal{L}=-\Lambda$ on cosmological scales, 
the background is identical to the background of 
$\Lambda$CDM.
In general this is always true. In fact, if we consider Eq.~(\ref{eq_phi})
or, equivalently, the continuity equations $(d \rho/d N)=-3(p+\rho)$,
and if we impose that $p=-\Lambda$, we easily get 
\begin{equation}
\rho=\rho_{{\rm DM}}(a=1)~a^{-3}+\Lambda= \rho_{\rm DM}+ \rho_\Lambda \;,
\end{equation}
where $\rho_\Lambda$ behaves like a cosmological constant ``dark energy'' 
component 
($\rho_\Lambda = {\rm const.}$) and 
$\rho_{\rm DM}$ behaves like a ``dark matter'' component 
($\rho_{\rm DM}\propto a^{-3}$).  
This result implies that we can think the stress tensor of our scalar field 
as being made of two components: one behaving like a
pressure-less fluid, and the other having negative pressure. 
In this way the integration constant $\rho_{{\rm DM}}(a=1)$
can be interpreted as the ``dark matter'' component today; consequently, 
$\Omega_m(0)=\rho_{{\rm DM}}(a=1)/(3H^2(a=1))$ and 
$\Omega_\Lambda(0)=\Lambda/(3H^2(a=1))$ are the 
density parameters of ``dark matter''and ``dark energy'' today.

Let us now describe the procedure that we will use in order to find UDM models with a small  speed of sound. 
By imposing the condition  $\mathcal{L}(X,\varphi)=-\Lambda$,
we constrain the solution of the equation of motion to live
on a particular manifold $\mathcal{M}_\Lambda$ embedded 
in the four dimensional space-time. This enables us 
to define $\varphi$ as a function of $X$ along 
the classical trajectories, i.e.
$\varphi=\mathcal{L}^{-1}(X,\Lambda)\big|_{\mathcal{M}_\Lambda}$.
Notice that therefore, by using Eq.(\ref{eq_phi_fg-v}) and 
imposing the constraint $p=-\Lambda$, i.e. 
$V(\varphi)=f(\varphi)g(X)+\Lambda$, we
can obtain the following general solution of the equation of motion
on the manifold $\mathcal{M}_\Lambda$
\begin{equation}
\label{mastereq}
2X\frac{\partial g(X)}{\partial X}f(\varphi(X))=\Lambda ~\nu ~a^{-3} \;,
\end{equation}
where $\nu\equiv \Omega_m(0)/\Omega_\Lambda(0)\,$ .
Here we have constrained the pressure to be $p=-\Lambda$.
In Section \ref{GUDM} we will describe an even more general technique 
to reconstruct
UDM models where the pressure is a free function of the scale factor $a$.

If we define the function $g(X)$, we immediately know the functional form 
of $c_s^2$ with respect to $X$ (see Eq.~(\ref{c_s^2_fg-v})). 
Therefore, if we have a Lagrangian
of the type $\mathcal{L}=f(\varphi)g(X)$ or $\mathcal{L}=g(X)-V(\varphi)$,
we are unable to decide 
the evolution of $c_s^2(X)$ along the solutions of the equation of motion
\cite{Bertacca:2007ux} because, once $g(X)$ is chosen, the 
constraint ${\cal L}=-\Lambda$ fixes immediateley the value of 
$f(\varphi)$ ($V(\varphi)$). 
On the contrary, in the case of Eq.~(\ref{Lagrangian}), we can do it 
through the function $f(\varphi(X))$. In fact, by properly defining
the value of $f(\varphi(X))$ and using Eq.(\ref{eq_phi_fg-v}), 
we are able to fix the slope of $X$ and, consequently (through $g(X)$),  
the trend of $c_s^2(X)$ as a function of the scale factor $a$.  

Finally, we want to emphasize that this approach is only a method
to reconstruct the explicit form of the Lagrangian (\ref{Lagrangian}), namely 
to separate the two variables $X$ and $\varphi$ into the
functions $g$, $f$ and $V$.

Now we give some examples where we apply this prescription.
In the following subsection, we consider the
explicit solutions when we assume a kinetic term of Born-Infeld type 
\cite{prc,abramo1,abramo2,Alishahiha:2004eh,stability1,stability2}.
Other examples (where we have the kinetic term $g(X)$
of the Scherrer model \cite{Scherrer:2004au} or where
we consider the generalized Scherrer solutions \cite{Bertacca:2007ux}) 
are reported in Appendix A.

\subsection{Lagrangians with Born-Infeld type kinetic term}
\label{BIL0}
Let us consider the following kinetic term
\begin{equation}
g(X)=-\sqrt{1-2X/M^4} \;, 
\end{equation}
with $M$ a suitable mass scale.
We get
\begin{equation}
\label{mastereq-BI}
\frac{2X/M^4}{\sqrt{1-2X/M^4}}f(\varphi(X))=\Lambda ~\nu ~a^{-3}\;,
\end{equation}
and
\begin{equation}
\label{c_s-BI}
c_s^2(X)=1-2X/M^4\;.
\end{equation}

At this point it is useful to provide two explicit examples where we show 
the power of this approach. In the next section, we give the example {\it par excellence}: 
a Lagrangian where the sound speed can be small. 
It is important to emphasize that the models described here and 
in the next section satisfy
the weak energy conditions $\rho\ge 0$ and $p+\rho\ge 0$. 

\begin{itemize} 
\item {\it Example~1)} \\
By defining $f$ as 
\begin{equation}
\label{mastereq-B_1}
f(\varphi(X))=\Lambda \frac{\left(1-2X/M^4\right)^{3/2}}
{\left(2X/M^4\right)^2}\;,
\end{equation}
we get
\begin{equation}
X(a)=\frac{M^4/2}{1+\nu a^{-3}}\;.
\end{equation}
In order to obtain an expression for $\varphi(a)$, we use
Eq.~(\ref{eq_u1}) and find 
\begin{equation}
\varphi(a)=
\left(\frac{M^2}{3\Lambda}\right)^{1/2}
\ln\left(\frac{1+\nu a^{-3}}{\nu a^{-3}}\right)\;.
\end{equation}
Now using Eq.~(\ref{mastereq-B_1}) and 
our initial ansatz $p=-\Lambda$ we obtain 
\begin{equation}
f(\varphi)=\frac{\Lambda}{4}\frac{\sinh\left[-
\left(\frac{3\Lambda}{4M^4}\right)^{1/2}\varphi\right]+\cosh\left[-
\left(\frac{3\Lambda}{4M^4}\right)^{1/2}\varphi\right]}
{\left\{\sinh\left[-\left(\frac{3\Lambda}{4M^4}\right)^{1/2}
\varphi\right]\right\}^2}
\end{equation}
and 
\begin{equation}
V(\varphi)=\frac{\Lambda}{4}\frac{\sinh\left[
\left(\frac{3\Lambda}{M^4}\right)^{1/2}\varphi\right]+\cosh\left[
\left(\frac{3\Lambda}{M^4}\right)^{1/2}\varphi\right]-2}
{\left\{\sinh\left[-\left(\frac{3\Lambda}{4M^4}\right)^{1/2}
\varphi\right]\right\}^2}\;.
\end{equation}

We can immediately see that $d X/dN>0$. 
Therefore, when $a \rightarrow 0$ we have $c_s^2 \rightarrow 1$, 
whereas when $a \rightarrow \infty$,  $c_s^2 \rightarrow 0$.
In other words, this model describes a unified fluid of 
dark matter and cosmological constant which is unavoidably in conflict 
with cosmological structure formation. 

\item {\it Example~2)} \\
Let us define 
\begin{equation}
\label{mastereq-B_2}
f(\varphi(X))= \frac{\Lambda} {\left(1-2X/M^4\right)^{1/2}}\;;
\end{equation}
then we get
\begin{equation}
X(a) =\frac{M^4}{2}\frac{\nu a^{-3}}{1 + \nu a^{-3}} \;.
\end{equation}
Following the same procedure adopted in the previous example,
we obtain 
\begin{equation}
\label{phi-B_2}
\varphi(a) =\frac{2 M^2}{\sqrt{3\Lambda}} 
\left\{ \arctan \left[\left( \nu  a^{-3}\right)^{-1/2}\right] 
- \frac{\pi}{2} \right\} \;.
\end{equation}
We immediately recover the same model studied in Ref.~\cite{Bertacca:2007ux}: 
\begin{equation}
f(\varphi)=\frac{\Lambda}{\left\vert\cos 
\left[ \left(\frac{3\Lambda}{4M^4}\right)^{1/2}\varphi \right] 
\right\vert}\;, \quad \quad \quad V(\varphi)=0 \;.
\end{equation}
In this case, the $c_s^2$ dependence on the scale factor $a$ is exactly 
opposite to the previous example: we have $c_s^2 \rightarrow 0$ 
when $a \rightarrow 0$, and $c_s^2 \rightarrow 1$ 
when $a \rightarrow \infty$. 
In this model, as explained in Ref.~\cite{Bertacca:2007cv}, 
the non-negligible value of the sound speed today gives a 
strong contribution to the ISW effect and produces an incorrect 
ratio between the first peak and the plateau of
the CMB anisotropy power-spectrum 
$l(l+1)C_l/(2\pi)$.  
In \ref{UDMradiation}, 
we study the kinematic behavior of this UDM fluid during the 
radiation-dominated epoch and we investigate for what values of $\varphi$
the kinetic term $X$ generates an appropriate basin of attraction. 
\end{itemize}

\section{UDM models with Born-Infeld type kinetic term and a low speed of sound}
\label{BIL}

Following the study of the second
example of the previous section, we now improve the dependence of $c_s^2$ on $a$
when $a \rightarrow \infty$. Let us consider for $f$ the 
following definition 
\begin{equation}
\label{f-B_3}
f(\varphi(X))= \frac{\Lambda}{\mu} ~\frac{2X/M^4-h}
{2X/M^4\left(1-2X/M^4\right)^{1/2}}\;,
\end{equation}
where $h$ and $\mu$ are appropriate positive constants. 
Moreover, we impose that $h<1$.
Thus we get
\begin{equation}
\label{X-B_3}
\fl
X(a) =\frac{M^4}{2}\frac{h + \mu \nu a^{-3}}{1 + \mu \nu a^{-3}} 
\quad {\rm or} \quad 
\left(\frac{d \varphi}{d N}\right)^2 = \frac{3 M^4}{\Lambda} 
\;\frac{h + \mu \nu a^{-3}}
{\left(1 + \nu a^{-3}\right) \left(1 + \mu \nu a^{-3}\right)}\;,
\end{equation}
and, for $c_s^2$, we obtain the following relation  
\begin{equation}
c_s^2(a)=\frac{1-h }{1 + \mu \nu a^{-3}}\;.
\end{equation}
Therefore, with the definition (\ref{f-B_3}) and using 
the freedom in choosing the value of $h$, 
we can shift the value of $c_s^2$ for  $a \rightarrow \infty$.
Specifically, $h=1-c_{\infty}^2$ where 
$c_{\infty}=c_s(a \rightarrow \infty)$.
At this point, by considering the case where $h=\mu$ (which makes the 
equation analytically integrable), we can
immediately obtain the trajectory $\varphi(a)$, namely 
\begin{equation}
\label{phi-B_3}
\varphi(a) =\left(\frac{4 h M^4}{3\Lambda} \right)^{1/2}
{\rm arc}\sinh\left( \nu h a^{-3}\right)^{-1/2} \;.
\end{equation}
Finally, we obtain 
\begin{equation}
\fl
f(\varphi)=\frac{\Lambda(1-h)^{1/2}}{h}
\frac{\cosh\left[\left(\frac{3\Lambda}{4hM^4}\right)^{1/2}
\varphi\right]}{\sinh\left[\left(\frac{3\Lambda}{4hM^4}\right)^{1/2}
\varphi\right]\left\{1+h\sinh^2\left[\left(\frac{3\Lambda}{4hM^4}\right)^{1/2}
\varphi\right]\right\}} \;,
\end{equation}
and
\begin{equation}
V(\varphi) = \frac{\Lambda}{h}\frac{\left\{h^2\sinh^2
\left[\left(\frac{3\Lambda}{4hM^4}\right)^{1/2}
\varphi\right]+2h-1\right\}}{1+h\sinh^2
\left[\left(\frac{3\Lambda}{4hM^4}\right)^{1/2}
\varphi\right]} \;.
\end{equation}
This result implies that in the early universe 
$\sqrt{3\Lambda/(4hM^4)}~ \varphi \ll 1$ and 
$2X/M^4 \approx 1$, and we obtain 
\begin{eqnarray}
\fl f(\varphi)\approx \left(\frac{4hM^4}{3\Lambda}\right)^{1/2}
\frac{\Lambda\sqrt{1-h}}{h} 
~\frac{1}{\varphi}~\propto ~a^{3/2}\;,
\quad \quad \quad \left|g(X)\right|= \sqrt{1-2X/\Lambda}
~\propto ~ a^{-3/2}\;,\nonumber \\ 
\fl |V(\varphi)|~ \longrightarrow ~\left|\frac{\Lambda(2h-1)}{h}\right| ~\ll~  
f(\varphi)\left(2X\frac{\partial g(X)}{\partial X}-g(X)\right)
~\propto ~ a^{-3}\;.
\end{eqnarray}
In other words, we find, for $ f(\varphi)$ and $ g(X)$,
a behaviour similar to that of {\it Example 2)}, as also 
obtained in Ref.~\cite{Bertacca:2007ux}
for a UDM Lagrangian of the type $\mathcal{L}= f(\varphi) g(X)$.\\

When $a \rightarrow \infty$, we have $\varphi \rightarrow \infty$ and
$2X/M^4 \rightarrow h$. Therefore
$$
f(\varphi) g(X) \longrightarrow 0\;, \quad \quad \quad  
V(\varphi)~ \longrightarrow \Lambda \;,
$$
that is, for $a \rightarrow \infty$, the dark fluid of this UDM model will 
converge to a Cosmological Constant.

Because the dark fluids described by this Lagrangian and the Lagrangian 
defined in {\it Example 2)}
behave similarly at early times, we conclude that the relative amounts of 
DM and DE that characterize
the present universe are fully determined by the value of
$\varphi (a\sim 0)$. In other words, to reproduce the present
universe, one has to tune the value of $f(\varphi)$ in the early universe.
However, as we analytically show in \ref{UDMradiation}, 
once the initial value of 
$\varphi$ is fixed, there is still a large basin of attraction in terms of the 
initial value of $d\varphi / dt$, which can take any value such that 
$2X/M^4 \ll 1$. Moreover, in \ref{UDMradiation}, we analytically investigate
the kinematic behavior of this UDM fluid during the
radiation-dominated epoch. \\

\begin{figure}
\centerline{\includegraphics[width = 6in,keepaspectratio=true]{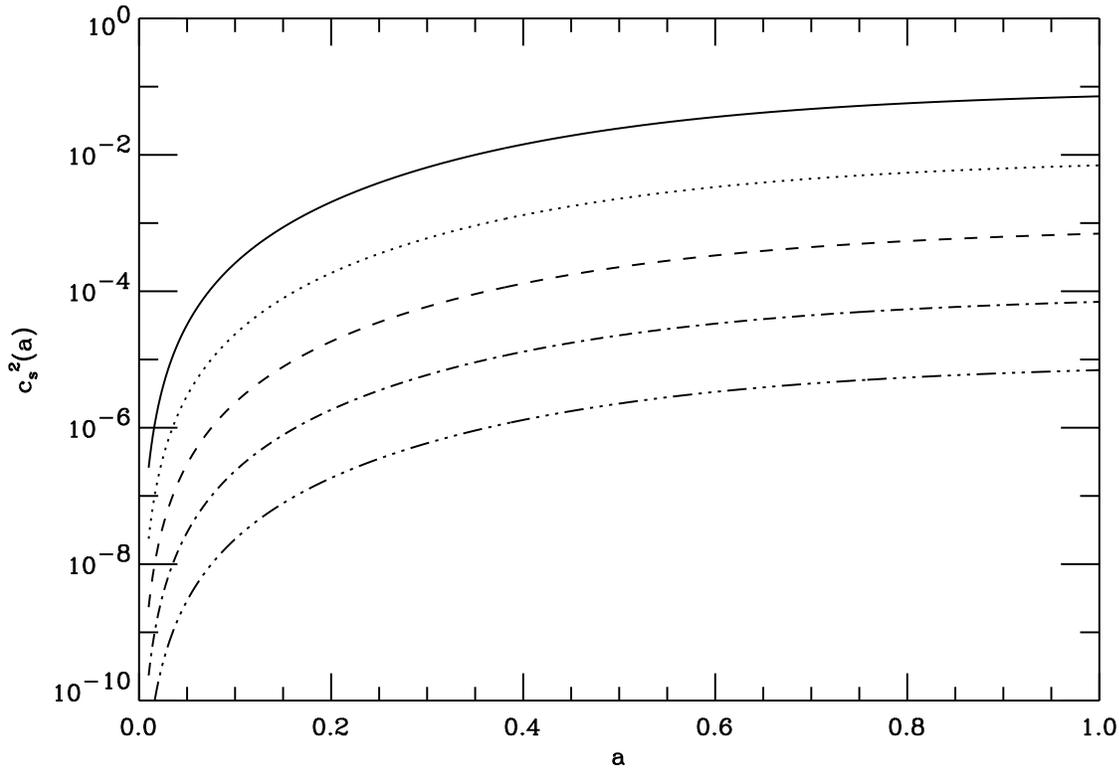}}
\caption{Sound speed velocity vs. the scale factor $a$ for 
different values of $c_{\infty}=10^{-1}, 10^{-2}, 10^{-3}, 10^{-4}, 10^{-5}$,
from top to bottom.}
\label{cs2}
\end{figure}

Finally, we can conclude that, once it is constrained to yield
the same  background evolution as $\Lambda$CDM and 
we set an appropriate value of $c_{\infty}$, 
this UDM model provides a sound speed small
enough that i) the dark fluid can cluster and ii) the Integrated Sachs-Wolfe contribution 
to the CMB anisotropies is compatible with observations.
Figure \ref{cs2} shows an example of the dependence of $c^2_s$ on $a$
for different values of $c_{\infty}$. 

\section{Prescription for UDM Models with a generic kinetic term}
\label{UDMg(X)-generic}
We now describe a general prescription 
to obtain a collection of models
that reproduce a background similar to $\Lambda$CDM 
and have a suitable sound speed. 
Some comments about the master equation (\ref{mastereq}) are first necessary.
The relation (\ref{mastereq}) enables to determine a connection
between the scalar factor $a$ and the kinetic term $X$ on the 
manifold $\mathcal{M}_\Lambda$ and therefore a mapping between 
the cosmic time and the manifold $\mathcal{M}_\Lambda$.

Now it is easy to see that
the LHS of Eq.~(\ref{mastereq}), seen as a single function of $X$,
must have at least a vertical asymptote and a zero, 
and the function must be continuous between the two. 
In particular, when $X$ is
near the vertical asymptote the universe approaches the cosmological 
constant regime,
whereas when $X$ is close to the zero of the function,
the dark fluid behaves like dark matter. 
Therefore, if we define
\begin{equation}
f(\varphi(X))=\frac{\mathcal{F}(X)}{2X(\partial g(X)/\partial X)}
\end{equation}
where, for example,
\begin{equation}
\mathcal{F}(X)=\frac{1}{\mu}\frac{X_f-X}{X-X_i}\;,
\end{equation}
(where $\mu$ is an appropriate positive constant)
the value of $X_f$ and $X_i$ are the zero and the asymptote mentioned above,
namely, when $a \rightarrow 0$ we have $X \rightarrow X_i$ and
when $a \rightarrow \infty$ we have $X \rightarrow X_f$.
Moreover, if $X_f>X_i$ we have $dX/dN>0$, whereas if $X_f<X_i$ 
we have $dX/dN<0$.
In other words, according to Eq.(\ref{mastereq}), 
\begin{equation}
X(a)=X_f \frac{1+(X_i/X_f)\Lambda \mu \nu ~a^{-3}}
{1+\Lambda \mu \nu ~a^{-3}}\;.
\end{equation}
Let us emphasize that the values of $X_i$ and $X_f$ are very important
because they automatically set the range of values 
that the sound speed can assume at the various cosmic epochs.

Let us finally make another important comment. One can use
this reconstruction of the UDM model in the opposite way.
In fact, by imposing a cosmological background
identical to $\Lambda$CDM, the observed CMB power spectrum, 
and the observed evolution of cosmic structures, we can derive the 
evolution of the sound speed $c_s^2$ vs. cosmic time.
In this case, by assuming an appropriate kinetic term $g(X)$ through Eq.~(\ref{c_s^2_fg-v}), 
we can derive $X(a)$ and, consequently,
$\varphi(a)$ and $X(a(\varphi))=X(\varphi)$. Therefore, by using the
relations (\ref{mastereq}) and $V(\varphi)=f(\varphi)g(X)+\Lambda$,
we can determine the functional form of $f(\varphi)$ and $V(\varphi)$.

%%%%%%%%%%%%%%%%%%%%%%%%%%%%%%%%%%%%%%%%%%%%%%%%%%%%%%%%%%%%%%%%%%%%%%%%%%

\section{A Particular Equivalence Class of UDM models}
\label{Class-eq}

In this section we investigate different UDM Lagrangians that have the same 
equation of state parameter $w$ and speed of sound $c_s$.
We show that a class of equivalent Lagrangians 
that have similar kinematical properties exists.
\ref{GeneralClass-eq} gives the most general derivation of this class.
Here, we describe a restricted class to emphasize the general procedure.
Let us begin with the Lagrangian  
\begin{equation}
\mathcal{L}=\mathcal{L}(h(\varphi)X,\varphi)\;,
\end{equation}
with $h(\varphi)>0$. It is very easy to show that, if $h(\varphi)\neq 0$, 
a field-redefinition $\phi \longrightarrow \psi$ exists
such that 
\begin{equation}
Y=\frac{\dot{\psi}^2}{2}=h(\varphi)X\quad\quad {\rm and}\quad\quad \psi=\pm
\int^{\varphi}[h(\tilde{\varphi})^{1/2}d\tilde{\varphi}]+K \;,
\end{equation}
where $K$ is an appropriate integration constant. Without any 
loss of generality, consider the case with the $+$ sign in front of the 
integral above. 

By performing this coordinate transformation, the Lagrangian becomes
\begin{equation}
\mathcal{L}(h(\varphi)X,\varphi)=\mathcal{L}(Y,\psi) \;,
\end{equation}
and the equation of motion (\ref{eq_phi}) becomes
\begin{equation}
 \label{eq_psi}
\left(\frac{\partial p}{\partial Y} 
+2Y\frac{\partial^2 p}{\partial Y^2}\right)\ddot\psi
+\frac{\partial p}{\partial Y}(3H\dot\psi)+
\frac{\partial^2 p}{\partial \psi \partial Y}\dot\varphi^2
-\frac{\partial p}{\partial \psi}=0 \;. 
\end{equation}
The most important property of this transformation is that 
the dependences of the equation of state and the effective speed of sound 
on the scale factor $a$ remain the same 
\footnote{It has been shown that models with the pure kinetic Lagrangian 
$\mathcal{L}(Y)$ 
(see for example Ref.~\cite{Bertacca:2007ux}) can be described as an 
adiabatic perfect fluid 
with pressure $p$ uniquely determined by the energy density, 
because both the pressure and the energy density depend on a single degree of 
freedom, the kinetic term $Y$. Thus, through this transformation, we can 
extend the adiabatic fluid Lagrangians studied in Ref.~\cite{Bertacca:2007ux} 
to a more general class of equivalent models.}.
In terms of the new variables $\psi$ 
and $Y$, one obviously has 
\begin{equation}
\label{w-psi}
w(\psi,Y) = \frac{p}{2Y \frac{\partial p}{\partial Y} - p} \;, 
\end{equation} 
and
\begin{equation}
\label{cs-psi}
c_s^2(\psi,Y) =  \frac{(\partial p /\partial Y)}
{(\partial \rho/\partial Y)} = 
\frac{\frac{\partial p }{\partial Y}}{\frac{\partial p}
{\partial Y}+ 2Y\frac{\partial^2 p}{\partial Y^2}} \;. 
\end{equation}

Is is easy to see that the transformations to the 
new variables used in Ref.~\cite{Tsujikawa:2004dp} to study scaling 
solutions are a particular case of this general prescription 
(see also \ref{GeneralClass-eq}).

Obviously, we can make the reverse reasoning (see 
\ref{GeneralClass-eq}): namely, by
starting from the Lagrangian dependent on $\psi$ and $Y$, we can obtain 
several Lagrangians of type $\mathcal{L}(\mathcal{R}(\theta)Z,\theta)$
with $Y=\mathcal{R}(\theta)Z$, where 
\begin{equation}
Z=\dot{\theta}^2/2\quad\quad {\rm and}\quad\quad \theta=
\int^{\psi}[\mathcal{R}(\tilde{\psi})^{-1/2}d\tilde{\psi}]+\mathcal{K} \;,
\end{equation}
where $\mathcal{K}$ is an appropriate integration constant and
where we have used the fact that $\mathcal{R}(\theta)$ becomes a function
of $\psi$, $\mathcal{R}[\theta(\psi)]$, thanks to the above coordinate 
transformation. 
Therefore, by considering the models obtained in the previous section and in 
\ref{otherUDM}, we can get different Lagrangians that have the same $w$ 
and $c_s^2$ evolution but have different kinematical properties.  
For instance, if we start from Eq.~(\ref{Lagrangian}), we get  
\begin{equation}
\label{LXphi-Ztheta}
\mathcal{L}=f(\varphi)g(X)-V(\varphi) =f(\theta)g(\mathcal{R}(\theta)Z)
- V(\theta)
\end{equation}
with 
$X=\dot{\varphi}^2/2=\mathcal{R}(\theta)Z=\mathcal{R}(\theta)
\dot{\theta}^2/2$ and for simplicity we write $f(\theta)\equiv f(\varphi(\theta))$ and $V(\theta)\equiv V(\varphi(\theta))$.

Now we describe some cases obtained starting from the model  
of Section \ref{BIL}~. In this way we can obtain Lagrangians
with kinetic term of Dirac-Born-Infeld type \cite{Alishahiha:2004eh}.
First of all, we consider an appropriate variable that
simplifies the functions $f(\varphi)$ and $V(\varphi)$. 
In fact, if $\mathcal{R}[\theta(\varphi)]^{-1/2} = 
\cosh{(\gamma \varphi)}$ where 
$\gamma=\left[(3\Lambda)/(4hM^4)\right]^{1/2}$, we get 
the following simplified Lagrangian
\begin{eqnarray}
\fl
\mathcal{L}(\theta,Z)=-\frac{\Lambda c_\infty}{1- c_\infty^2}
\frac{(1+\gamma^2\theta^2)^{1/2}}{\gamma 
\theta[1+ (1-c_\infty^2)\gamma^2\theta^2]}
\left[1-\frac{2Z/M^4}{(1+\gamma^2\theta^2)}\right]^{1/2}+\nonumber \\
-\frac{\Lambda}{1- c_\infty^2}\frac{(1-c_\infty^2)^2\gamma^2\theta^2+1-2c_\infty^2}
{[1+ (1-c_\infty^2)\gamma^2\theta^2]}\;.
\end{eqnarray}
Thus, by using the coordinate transformation, we obtain the following relations 
\begin{equation}
\theta(a)=\frac{1}{\gamma}\frac{1}{[(1-c_\infty^2)\nu a^{-3}]^{1/2}}\;,
\end{equation}
and
\begin{equation}
Z(a)=\frac{M^4}{2}\frac{1+\nu a^{-3}}{\nu a^{-3}}\;.
\end{equation}
Another possibility to obtain a simpler Lagrangian is to define 
$\mathcal{R}[\theta(\varphi)]^{-1/2} = 
2\cosh{(\gamma \varphi)}\sinh{(\gamma \varphi)}$.
In \ref{CoordTransform}, we give more examples of how coordinate transformations 
of Lagrangians with a Born-Infeld kinetic term can yield
Lagrangians with different kinematical properties.

%%%%%%%%%%%%%%%%%%%%%%%%%%%%%%%%%%%%%%%%%%%%%%%%%%%%%%%%%%%%%%%%%%%%%%%%%%%

\section{Generalized UDM Models}
\label{GUDM}

In this Section we consider several possible generalizations of the technique 
introduced in Section \ref{UDM}, with the aim of  
studying models where the background does not necessarily mimic
the $\Lambda$CDM background.
Finally, we want to emphasize that the Lagrangians
we obtain here can also be generalized by means of the field redefinition 
defined above and firther detailed in \ref{GeneralClass-eq}.
We can write the Lagrangian with two different simple approaches:\\
\begin{itemize} 
\item[1)]By choosing $p(N)$. Indeed we get
\begin{equation}
\fl
\frac{d\rho}{dN}+3\rho=-3p(N)\;,\quad\quad{\rm i.e.}
\quad\quad \rho(N)=e^{-3N}\left[-3\int^N\left(e^{3N'}p(N')dN'\right) + 
K\right]\;,
\end{equation}
where $K$ is an integration constant.
By imposing the condition $\mathcal{L}(X,\varphi)=p(N)$ along 
the classical trajectories, we obtain 
$\varphi=\mathcal{L}^{-1}(X(N),p(N))\big|_{\mathcal{M}_{p(N)}}$.
Thus, starting from a generic Lagrangian $\mathcal{L} = 
f(\varphi)g(X)-V(\varphi)$ we get
\begin{equation}
\label{gen2}
\fl
2X(N)\left[\frac{\partial g(X)}{\partial X}\right](N)f(\varphi(X,N)) = 
p(N)+e^{-3N}\left[-3\int^N\left(e^{3N'}p(N')dN'\right)+K\right]\;.
\end{equation}
For example, if $p=-\Lambda$, $K=\rho(a=1)$. The freedom provided by
the choice of $K$ is particularly relevant. In fact,
by setting $K=0$, we can remove the term $\rho \propto a^{-3}$. Alternatively, 
when $K\ne 0$, we always have a term that behaves like presseure-less matter.
We thus show that the single fluid of UDM models can mimic
not only a cosmological constant but also any quintessence fluid.

Thus, using  Eq.~(\ref{gen2}) and by following the argument described 
in Section \ref{UDM}, we can get the relations
$X\equiv\mathcal{G}_p(N)$, and consequently 
\begin{eqnarray}
\fl \varphi & \equiv & \mathcal{Q}_p (N) = \varphi_0 \nonumber \\
\fl & \pm & 
\int^N\left\{\mathcal{G}_p (N')^{1/2}
\left[-3e^{-3N}\int^N\left(e^{3N'}p(N')dN'\right) + 
Ke^{-3N}\right]^{-1/2}dN'\right\}\;.
\end{eqnarray}
Therefore, with the functions $\mathcal{G}_p (N)$ and $\mathcal{Q}_p (N)$, 
we can write 
$f(X,N)=f(\mathcal{G}_p(N),N)=
f(\mathcal{G}_p(\mathcal{Q}_p^{-1}(\varphi)),\mathcal{Q}_p^{-1}(\varphi)) = 
f(\varphi)$.
Thus, by starting from a Lagrangian whose behavior is given  
by $p(N)$, the speed of sound is determined by the appropriate 
choice of $g(X)$. 

\item[2)]By choosing the equation of state $w(N)$. Indeed
\begin{equation}
\rho(N)=\rho_{0}e^{-3\int^N(w(N')+1)dN'}\;,
\end{equation}
where $\rho_0$ is a positive integration constant, and
\begin{equation}
p(N)=\rho_{0}w(N)e^{-3\int^N(w(N')+1)dN'}\;.
\end{equation}
Therefore, still by imposing the condition $\mathcal{L}(X,\varphi) = 
p[w(N),N]$ along 
the classical trajectories, i.e.
$\varphi=\mathcal{L}^{-1}[X(N),p(w(N),N)]\big|_{\mathcal{M}_{w(N)}}$, we get
\begin{equation}
\label{gen3}
2X\frac{\partial g(X)}{\partial X}f(X,N)=\rho_0[w(N)+1]
e^{-3\int^N(w(N')+1)dN'}\;.
\end{equation}
Therefore, on the classical trajectory we can impose, by using $w(N)$, 
a suitable function $p(N)$ and thus the function $\rho(N)$.
The master equation Eq.~(\ref{gen3}) generalizes Eq.~(\ref{mastereq}).
Also in this case, by  Eq.~(\ref{gen3}) and by following the 
argument described in Section \ref{UDM}, we can get the relations
$X\equiv\mathcal{G}_w(N)$, and consequently 
\begin{equation}
\fl
\varphi\equiv\mathcal{Q}_w(N)=\pm \int^N\left\{\mathcal{G}_w(N')^{1/2}
\left[\rho_{0}e^{-3\int^{N'}(w(N'')+1)dN''}\right]^{-1/2}dN'\right\} + 
\varphi_0\;.
\end{equation}
Thus, with the functions $\mathcal{G}_w(N)$ and $\mathcal{Q}_w(N)$, 
we can write 
$f(X,N)=f(\mathcal{G}_w(N),N)=
f(\mathcal{G}_w(\mathcal{Q}_w^{-1}(\varphi)),\mathcal{Q}_w^{-1}(\varphi))
= f(\varphi)$.
Then we can find a Lagrangian whose behavior is determined  
by $w(N)$ and whose speed of sound is determined by the appropriate 
choice of $g(X)$. 
\end{itemize}

Finally, we conclude that the $p(N)$ constraint on the 
equation of motion is actually a  weaker condition than the $w(N)$ 
constraint. The larger freedom that the $p(N)$ constraint provides
naturally yields an additive term in the energy density that decays
like $a^{-3}$, i.e. like a matter term 
in the homogeneous background. 
Let us emphasize that this important result is a natural consequence of the 
$p(N)$ constraint and is not imposed {\it a priori}. 
 
\section{Conclusions}

A general severe problem of many UDM models considered so far 
is that their large effective speed 
of sound causes a strong time evolution of the gravitational potential 
and generates an ISW effect much larger than current 
observational limits.
In this paper we have outlined a technique to reconstruct UDM models
such that the effective speed of sound is small enough
that these problems are removed and the scalar field can cluster.
 
We have also considered a more general class of UDM Lagrangians 
with a non-canonical kinetic term. Specifically, we have studied some 
invariance properties of general Lagrangians of the form 
$\mathcal{L} = \mathcal{L}(h(\varphi)X,\varphi)$ 
which allows to define different models 
whose equations of motion are dynamically equivalent. 

Finally, we have studied a possible way to generalize UDM models 
that can mimic a fluid of dark matter and quintessence-like dark energy. 

The Lagrangians that we obtained appear rather contrived. Indeed, 
these models should be understood as examples which show that the 
mechanism itself can work. These models can however help to search for
physically motivated models with the desired properties. 
Moreover many previous work attempted to solve this problem by a severe
fine-tuning of the parameters appearing in the Lagrangian.
This drawback does not belong to our models. 

In future work, we will consider models with Lagrangians 
$\mathcal{L}=\mathcal{L}(X,\varphi)$
to estimate astrophysical observables, like the cross-correlation of 
CMB anisotropies and large-scale structure or the weak lensing shear 
signal power-spectrum.

%%%%%%%%%%%%%%%%%%%%%%%%%%%%%%%%%%%%%%%%%%%%%%%%%%%%%%%%%%%%%%%%%%%%%%%%%%%%%%

\section*{Acknowledgments}
We thank Alan Heavens, Pier Stefano Corasaniti and Luca Amendola for 
useful discussion.
Support from the PRIN2006 grant ``Costituenti fondamentali dell'Universo'' of
the Italian Ministry of University and Scientific Research and from the 
INFN grant PD51 is gratefully acknowledged.

%%%%%%%%%%%%%%%%%%%%%%%%%%%%%%%%%%%%%%%%%%%%%%%%%%%%%%%%%%%%%%%%%%%%%%%%%%
%APPENDIX
%%%%%%%%%%%%%%%%%%%%%%%%%%%%%%%%%%%%%%%%%%%%%%%%%%%%%%%%%%%%%%%%%%%%%%%%%%  
\vskip 1cm
\appendix
\setcounter{equation}{0}
\def\theequation{A.\arabic{equation}}
\vskip 0.2cm
\section{Other UDM model examples}
\label{otherUDM}
In this Appendix we will give further examples of UDM models
with small sound speed $c_s$. 

\subsection{Others possible Lagrangians with a kinetic term of 
Born-Infeld type}

Let us define $f(\varphi(X))$ in the following way
\begin{equation}
f(\varphi(X))= \frac{\Lambda}{\mu} ~\frac{\left(1-2X/M^4\right)^{3/2}}
{2X/M^4\left(2X/M^4-h\right)}\;,
\end{equation}
where $0<h<1$ and $\mu>0$.
Therefore, with Eq.~(\ref{mastereq-BI}), we obtain
\begin{equation}
\label{X-B_4}
\fl X(a) =\frac{M^4}{2}\frac{1 + h \mu \nu a^{-3}}{1 + \mu \nu a^{-3}} 
\quad  \quad {\rm or} \quad  \quad
\left(\frac{d \varphi}{d N}\right)^2 = \frac{3 M^4}{\Lambda} 
\;\frac{1 + h \mu \nu a^{-3}}
{\left(1 + \nu a^{-3}\right) \left(1 + \mu \nu a^{-3}\right)}\;.
\end{equation}
As it is easy to see, this Lagrangian has opposite properties 
to those of the model in Section \ref{BIL}. 
Indeed, here we have $d X/ d N > 0$ and, consequently, 
$2X/ M^4\to h$ at early times and approaches $1$ 
when $a \rightarrow \infty$. Therefore, using Eq.~(\ref{c_s-BI}),
we can conclude that $c_s^2\to 1-h$ when $a \rightarrow 0$ 
and zero when $a \rightarrow \infty$. 

A possible simple analytical solution can be obtained if we define $\mu=1/h$.
In fact, in this case we get 
\begin{equation}
\label{phi-B_4}
\varphi(a) =\left(\frac{4  M^4}{3\Lambda} \right)^{1/2}
{\rm arc}\sinh\left( \frac{\nu}{h} a^{-3}\right)^{-1/2} \;.
\end{equation}
This gives
\begin{equation}
\fl
f(\varphi)=\Lambda h \sqrt{1-h}
\frac{\cosh\left[\left(\frac{3\Lambda}{4M^4}\right)^{1/2}
\varphi\right]}{\sinh^2\left[\left(\frac{3\Lambda}{4M^4}\right)^{1/2}
\varphi\right]\left\{h+\sinh^2\left[\left(\frac{3\Lambda}{4M^4}\right)^{1/2}
\varphi\right]\right\}} \;,
\end{equation}
and
\begin{equation}
\fl
V(\varphi)=\Lambda
\frac{\left\{\sinh^4\left[\left(\frac{3\Lambda}{4M^4}\right)^{1/2} 
\varphi\right]+h \sinh^2\left[\left(\frac{3\Lambda}{4M^4}\right)^{1/2}
\varphi\right]-h(1-h)\right\}}
{\sinh^2\left[\left(\frac{3\Lambda}{4M^4}\right)^{1/2}
\varphi\right]\left\{h+\sinh^2\left[\left(\frac{3\Lambda}{4M^4}\right)^{1/2}
\varphi\right]\right\}} \;. 
\end{equation}
One can see that the speed of sound depends on the scale factor $a$ 
as follows 
\begin{equation}
c_s^2=(1-h)\frac{(\nu/h) a^{-3}}{1 +(\nu/h)a^{-3}}\;.
\end{equation}

Finally, one can generalize these models by imposing that the sound speed 
is zero neither when $a \rightarrow 0$  nor when  $a \rightarrow \infty$.
Consider the following relation 
\begin{equation}
f(\varphi(X))= \frac{\Lambda}{\mu} ~\frac{\left(1-2X/M^4\right)^{1/2}
(h_\infty-2X/M^4)}{2X/M^4\left(2X/M^4-h_0 \right)}\;,
\end{equation}
where $0<h_0<1$,~$0<h_\infty<1$  and $\mu>0$. Now $c_s^2\to 1-h_0$
in the early universe and $c_s^2\to 1-h_\infty$ when  $a \rightarrow \infty$.
In fact, 
\begin{equation}
c_s^2(a)=\frac{(1-h_\infty)+ (1-h_0) \mu \nu a^{-3} }{1 + \mu \nu a^{-3}}\;.
\end{equation}
Then
\begin{equation}
\label{X-B_5}
\fl X(a) =\frac{M^4}{2}\frac{h_\infty + h_0 \mu \nu a^{-3}}
{1 + \mu \nu a^{-3}} 
\quad  \quad {\rm or} \quad  \quad
\left(\frac{d \varphi}{d N}\right)^2 = \frac{3 M^4}{\Lambda} 
\;\frac{ h_\infty + h_0 \mu \nu a^{-3}}
{\left(1 + \nu a^{-3}\right) \left(1 + \mu \nu a^{-3}\right)}\;.
\end{equation}
Therefore if $\mu=h_\infty/ h_0$ we obtain
\begin{equation}
\label{phi-B_5}
\varphi(a) =\left(\frac{4 h_\infty M^4}{3\Lambda} \right)^{1/2}
{\rm arc}\sinh\left( \frac{h_\infty}{h_0} \nu a^{-3}\right)^{-1/2} \;,
\end{equation}
which gives
\begin{eqnarray}
\fl
f(\varphi)=\Lambda\frac{h_0}{h_\infty}
\frac{\cosh\left[\left(\frac{3\Lambda}{4h_\infty M^4}\right)^{1/2}
\varphi\right]\left\{(1-h_0)+(1-h_\infty)\sinh^2 
\left[\left(\frac{3\Lambda}{4h_\infty M^4}\right)^{1/2}\varphi\right]
\right\}^{1/2}}{\sinh^2\left[\left(\frac{3\Lambda}{4h_\infty M^4}\right)^{1/2}
\varphi\right]\left\{h_0+h_\infty\sinh^2\left[\left(\frac{3\Lambda}{4h_\infty 
M^4}\right)^{1/2}
\varphi\right]\right\}} \;,\nonumber \\
\end{eqnarray}
and
\begin{eqnarray}
\fl
V(\varphi)=\frac{\Lambda}{h_\infty}\frac{h_\infty 
^2\sinh^4\left[\left(\frac{3\Lambda}{4h_\infty M^4}\right)^{1/2}
\varphi\right]+h_0(2h_\infty-1)\sinh^2\left[\left(\frac{3\Lambda}{4h_\infty 
M^4}\right)^{1/2}
\varphi\right]-h_0(1-h_0)}{\sinh^2\left[\left(\frac{3\Lambda}{4h_\infty 
M^4}\right)^{1/2}
\varphi\right]\left\{h_0+h_\infty\sinh^2\left[\left(\frac{3\Lambda}{4h_\infty 
M^4}\right)^{1/2}
\varphi\right]\right\}} \;.\nonumber \\
\end{eqnarray}
It easy to see that these relations can be used both when $h_0<h_\infty$ 
(i.e.~$dX/dN>0$) and when $h_\infty < h_0$ (i.e.~$dX/dN<0$). 

\subsection{Lagrangian with kinetic term of the generalized Scherrer 
solutions type}

Consider the following kinetic term \cite{Bertacca:2007ux}
\begin{equation}
\label{g_k-n}
g(X) = g_n (X/M^4 - \hat{\chi})^n\;.
\end{equation}
with $n > 1$ and with $\hat{\chi}>0$. In this case the sound speed becomes
\begin{equation}
\label{c2_n}
c_s^2 = \frac{(X/M^4-\hat{\chi})}{2 (n-1) \hat{\chi} + (2n-1) 
(X/M^4-\hat{\chi})}\;.
\end{equation}
Moreover, if we set $\epsilon = [(X/M^4-\hat{\chi})/\hat{\chi}]  \ll 1$ 
we easily obtain
\begin{equation}
\label{c2_n-epsilon}
c_s^2 \simeq \frac{1}{2(n-1)} \epsilon \;. 
\end{equation}

Now Eq.~(\ref{mastereq}) takes the form
\begin{equation}
\label{mastereq-GS}
2n g_n (X/M^4) (X/M^4 - \hat{\chi})^{n-1} f(\varphi(X))=\Lambda \nu a^{-3}\;. 
\end{equation}
Below we provide an example of a UDM model where $dX/dN<0$ (i.e. for
$dc_s^2/dN<0$) and, finally,  we give an example that generalizes the
Lagrangians with the kinetic term of the generalized Scherrer solutions 
both for $dX/dN<0$ 
($dc_s^2/dN<0$) and for $dX/dN>0$ ($dc_s^2/dN<0$).
\subsubsection{$dX/dN<0$. }
Define
\begin{equation}
f(\varphi(X))= \frac{\Lambda}{\mu} 
~\frac{1}{2ng_n(X/M^4)(X/M^4 -\hat{\chi} )^{n-1}}
\frac{\left(X/M^4-\hat{\chi}\right)}{\left( \chi_i-X/M^4 \right)}\;,
\end{equation}
where $\chi_i>\hat{\chi}$. Then by Eq.~(\ref{mastereq-GS}) we get
\begin{eqnarray}
\label{X-GS1}
\fl X(a)/M^4 =\hat{\chi}\frac{1 +(\mu\chi_i/\hat{\chi})\nu a^{-3}}
{1 + \mu \nu a^{-3}} 
\quad \quad {\rm or} \quad\quad
\left(\frac{d \varphi}{d N}\right)^2 = \frac{6 M^4\hat{\chi}}{\Lambda} 
\;\frac{1 + (\mu\chi_i/\hat{\chi})\nu a^{-3}}
{\left(1 + \nu a^{-3}\right) \left(1 + \mu \nu a^{-3}\right)}\;.\nonumber \\
\end{eqnarray}
Now, if $\mu=\hat{\chi}/\chi_i$, we obtain the following relations
\begin{equation}
\label{phi-GS1}
\varphi(a) =\left(\frac{8M^4\hat{\chi}}{3\Lambda} \right)^{1/2}
{\rm arc}\sinh\left( \frac{\hat{\chi}}{\chi_i} \nu a^{-3}\right)^{-1/2} \;,
\end{equation}
\begin{equation}
\fl
f(\varphi)=\frac{\chi_i\Lambda}{2 n g_n \hat{\chi}(\chi_i- \hat{\chi})^{n-1}}
\frac{\cosh^{2n}\left[\left(\frac{3\Lambda}{8M^4\hat{\chi}}\right)^{1/2}
\varphi\right]}
{\sinh^2\left[\left(\frac{3\Lambda}{8M^4\hat{\chi}}\right)^{1/2}\varphi\right]
\left\{\chi_i+\hat{\chi}\sinh^2\left[\left(\frac{3\Lambda}{8M^4\hat{\chi}}
\right)^{1/2}
\varphi\right]\right\}} \;,
\end{equation}
\begin{eqnarray}
\fl
V(\varphi)=\frac{\Lambda}{2 n\hat{\chi}}
\frac{2 n\hat{\chi}^2\sinh^4\left[\left(\frac{3\Lambda}{8M^4\hat{\chi}}
\right)^{1/2}\varphi\right]+
2 n\hat{\chi}\chi_i\sinh^2\left[\left(\frac{3\Lambda}{8M^4\hat{\chi}}
\right)^{1/2}\varphi\right]+
\chi_i(\chi_i-\hat{\chi})}{\sinh^2\left[\left(\frac{3\Lambda}{8M^4\hat{\chi}}
\right)^{1/2}\varphi\right]
\left\{\chi_i+\hat{\chi}\sinh^2\left[\left(\frac{3\Lambda}{8M^4\hat{\chi}}
\right)^{1/2}
\varphi\right]\right\}}\;.
\end{eqnarray}
Then the sound speed is given by
\begin{equation}
c_s^2(a)=(\chi_i- \hat{\chi})\frac{(\hat{\chi}/\chi_i) \nu a^{-3}}
{2(n-1)\hat{\chi}+[(2n-1)\chi_i-\hat{\chi}](\hat{\chi}/\chi_i) \nu a^{-3}}\;.
\end{equation}
Therefore at early times 
$c_s^2\rightarrow(\chi_i- \hat{\chi})/ [(2n-1)\chi_i-\hat{\chi}]$
and when $a \rightarrow \infty$ we have $c_s^2\rightarrow 0$.
Moreover, if $\epsilon \ll 1 $ (provided $(\chi_i- \hat{\chi})/\hat{\chi} 
\ll  1$) the sound speed takes the form
\begin{equation}
c_s^2 \simeq \frac{1}{2(n-1)}
\frac{[(\chi_i- \hat{\chi})/\hat{\chi}]\nu a^{-3}}{1+(\hat{\chi}/\chi_i) 
\nu a^{-3}}\;.
\end{equation}

\subsubsection{General case.}
Consider the following relation
\begin{equation}
f(\varphi(X))= \frac{\Lambda}{\mu} 
~\frac{1}{2ng_n(X/M^4)(X/M^4 -\hat{\chi} )^{n-1}}
\frac{\left(X/M^4- \chi_f \right)}{\left( \chi_i-X/M^4 \right)}\;,
\end{equation}
where $\chi_i>\hat{\chi}>0$ and $\chi_f>\hat{\chi}$.
Then if $\mu=\chi_f/\chi_i$ we get the following relations
\begin{equation}
\label{X-GS}
X(a)/M^4 =\chi_f\frac{1 +\nu a^{-3}}
{1 + (\chi_f/\chi_i)\nu a^{-3}} \;,
\end{equation}
\begin{equation}
\label{phi-GS1}
\varphi(a) =\left(\frac{8M^4\chi_f}{3\Lambda} \right)^{1/2}
{\rm arc}\sinh\left( \frac{\chi_f}{\chi_i} \nu a^{-3}\right)^{-1/2} \;,
\end{equation}
\begin{eqnarray}
f(\varphi)=\frac{\chi_i\Lambda}{2 n g_n \chi_f}
\frac{\cosh^{2n}\left[\left(\frac{3\Lambda}{8M^4\chi_f}\right)^{1/2}
\varphi\right]}
{\sinh^2\left[\left(\frac{3\Lambda}{8M^4\chi_f}\right)^{1/2}\varphi\right]
\left\{\chi_i+\chi_f\sinh^2\left[\left(\frac{3\Lambda}{8M^4\chi_f}
\right)^{1/2}\varphi\right]\right\}} \nonumber \\
\left\{(\chi_i-\hat{\chi})+(\chi_f-\hat{\chi})\sinh^2
\left[\left(\frac{3\Lambda}
{8M^4\chi_f}\right)^{1/2}\varphi\right]\right\}^{1-n} \;,
\end{eqnarray}
\begin{eqnarray}
\fl
V(\varphi)=\frac{\Lambda}{2 n\chi_f}
\Bigg\{2 n \chi_f^2\sinh^4\left[\left(\frac{3\Lambda}{8M^4\chi_f}
\right)^{1/2}\varphi\right]+
\chi_i[(2 n+1)\chi_f-\hat{\chi}]\sinh^2\left[\left(
\frac{3\Lambda}{8M^4\chi_f}\right)^{1/2}\varphi\right]\nonumber\\
\fl +\chi_i(\chi_i-\hat{\chi})\Bigg\}
\sinh^{-2}\left[\left(\frac{3\Lambda}{8M^4\chi_f}\right)^{1/2}\varphi\right]
\left\{\chi_i+\chi_f\sinh^2\left[\left(\frac{3\Lambda}{8M^4\chi_f}\right)^{1/2}
\varphi\right]\right\}^{-1}\;.\nonumber \\
\end{eqnarray}
The sound speed is
\begin{equation}
c_s^2(a)=\frac{(\chi_f- \hat{\chi})+(\chi_i- \hat{\chi})(\chi_f/\chi_i) \nu 
a^{-3}}
{[(2n-1)\chi_f-\hat{\chi}]+[(2n-1)\chi_i-\hat{\chi}](\chi_f/\chi_i) 
\nu a^{-3}} \;,
\end{equation}
We can immediately see that at early times 
$c_s^2\rightarrow(\chi_i- \hat{\chi})/ [(2n-1)\chi_i-\hat{\chi}]$
and when $a \rightarrow \infty$ we have 
$c_s^2\rightarrow (\chi_f- \hat{\chi})/ [(2n-1)\chi_f-\hat{\chi}]$.
Therefore, with this Lagrangian, the sound speed can both grow and decrease, 
depending 
on the value taken by $\chi_i$ and $\chi_f$. 
Moreover, if $\epsilon \ll 1 $ we obtain
\begin{equation}
c_s^2 \simeq \frac{1}{2(n-1)}
\frac{[(\chi_f- \hat{\chi})/\hat{\chi}]+[(\chi_i- \hat{\chi})/
\hat{\chi}](\chi_f/\chi_i)\nu a^{-3}}
{1+(\chi_f/\chi_i) \nu a^{-3}}\;.
\end{equation}

%%%%%%%%%%%%%%%%%%%%%%%%%%%%%%%%%%%%%%%%%%%%%%%%%%%%%%%%%%%%%%%%%%%%%%%%%%%

\def\theequation{B.\arabic{equation}}
\vskip 0.2cm

\section{Study of UDM models when the yniverse is dominated by radiation}
\label{UDMradiation}
In general, when we do not neglect the radiation, the 
background evolution of the universe is completely characterized 
by the following equations 
\begin{equation}
\label{eq_u1R}
 H^2 =\frac{1}{3} (\rho + \rho_R)\, ,
\end{equation}
\begin{equation}
\label{eq_u2R}
\dot{H} = - \frac{1}{2}
(p + \rho + p_R + \rho_R)\, ,
\end{equation}
where $\rho_R$ and $p_R$
are the radiation energy density and pressure, respectively.

In this Appendix we consider the universe dominated by radiation
and we want to study analytically the behavior of
UDM models with kinetic term of Born-Infeld type.
In particular, we study the Lagrangian obtained in {\it Example~2)}
and in the model 
of Section \ref{BIL} ({\it Example~3)}). These Lagrangians have similar behavior at
early times; thus, because {\it Example~2)} is simpler than {\it Example~3)},
we investigate the former: the result will then also apply to {\it Example~3)}.
We proceed by defining some functions of the scale factor,
which make simples the study of the dynamics of these Lagrangians
when the universe is not dominated by the UDM field.

\subsection{Lagrangian of the type $\mathcal{L}(\varphi ,X)=f(\varphi)g(X)$}
\label{fg}
Let us introduce appropriate functions of the scale factor.
We write Eq.~(\ref{eq_phi}) as follows
\begin{eqnarray}
\label{background0}
\fl
\frac{1}{f}\frac{df}{dN}= \lambda(N,X) \nonumber \\
\fl
\left(\frac{\partial g}{\partial X} +  2X\frac{\partial^2 g}{\partial X^2}
\right) dX
+ \left[ 3 \left(2 X \frac{\partial g}{\partial X}\right) +
\lambda(N,X,f(N,X)) \left(2 X \frac{\partial g}{\partial X} - g \right) 
\right] dN = 0 \;. \nonumber\\
\end{eqnarray}
Eqs.~(\ref{background0}) define the quantity $\lambda$ as a generic 
function of $N$.

Now, in order to get a second function of the scale factor, we find 
the set of scalar field trajectories where 
the second of Eq.~(\ref{background0}) defines an exact differential form. 
To this aim, first of all we have to study the differential form $P(X,N)\; dX + Q(X,N)\; dN = 0$. 
One possible way to make it an exact differential form 
is to search for an integral factor $I$, which is an explicit 
function of $N$. 
In our situation $P(X,N)=P(X)$, thus $I(N)$ is 
\begin{equation}
\label{I}
\frac{dI}{I}= \frac{\frac{\partial Q(X,N)}{\partial X}}{P(X)} dN \; .
\end{equation}
In this case, we have to impose the integrability condition
\begin{equation}
\label{IC}
\frac{\partial Q(X,N)}{\partial X} = \alpha (N) P(X)
\end{equation}
so that  $I(N)= \exp{\int^N  dN' \alpha (N')}$ only depends on $N$.

Using the explicit expressions of $Q(X,N)$ and $P(X)$, the condition 
(\ref{IC}) becomes
\begin{equation}
\label{alpha_2}
 3 \frac{\partial \left(2 X \frac{\partial g}{\partial X}\right)}
{\partial X}+ \frac{\partial \lambda}{\partial X} 
\left(2 X \frac{\partial g}{\partial X} - g \right) + 
\left(\lambda - \alpha \right) \frac{\partial 
\left(2 X \frac{\partial g}{\partial X} - g\right)}{\partial X} = 0.
\end{equation}
It is easy to see that $\lambda - \alpha$ is a function of (at least) 
$X$; then, defining $G(X) \equiv \alpha - \lambda$, 
Eq.~(\ref{alpha_2}) becomes 
\begin{equation}
\label{alpha_3}
 3 \frac{\partial \left(2 X \frac{\partial g}{\partial X}\right)}
{\partial X}-\frac{\partial G}{\partial X} 
\left(2 X \frac{\partial g}{\partial X} - g \right) - 
G \frac{\partial \left(2 X \frac{\partial g}{\partial X} - 
g\right)}{\partial X} = 0
\end{equation}
which can be trivially integrated to give 
\begin{equation}
\label{alpha_5}
3 \left(2 X \frac{\partial g}{\partial X}\right) + K = 
G \left(2 X \frac{\partial g}{\partial X} - g \right) 
\end{equation}
with $K$ a generic constant. Without any loss of generality, we can 
set $K = 0$ so that 
\begin{equation}
\label{sigma}
\alpha - \lambda = G = 3 (w+1).
\end{equation}
By inserting Eq.~(\ref{alpha_5}) into the second of Eqs.~(\ref{background0}) 
we find 
\begin{equation}
\label{background1}
\left(\frac{\partial g}{\partial X}+2X\frac{\partial^2 g}
{\partial X^2}\right) dX +  \alpha (N) 
\left(2 X \frac{\partial g}{\partial X} -g \right) dN =0 \; .
\end{equation}
By multiplying both sides by $I(N)$, we finally obtain 
\begin{equation}
\label{general_solution}
\left(2 X \frac{\partial g}{\partial X} -g \right) = 
\bar{K} e^{ - \int^N dN' \alpha(N')}
\end{equation}
where $\bar{K} $ is a new integration constant.
Using the general equation (\ref{general_solution}),
we can express the energy density as 
\begin{equation}
\label{density_k}
\rho = \bar{K}  e^{ - \int^N dN' \left( \alpha(N')-\lambda(N')\right)} 
= \bar{K} e^{ - 3 \int^N dN' (w(N') + 1)}\;.
\end{equation}
If $\lambda \rightarrow 0$ and $\alpha \rightarrow 0$, 
$w\rightarrow -1 $ and $\bar{K} f\rightarrow {\rm const.}$ 
Therefore, the energy density $\rho $ tends to a constant 
$\rho_{0}$.\footnote{For $\lambda = 0$, the Lagrangian $\mathcal{L}$ (i.e. the pressure
$p$) depends 
only on $X$; in other words, we are obtaining the equations that describe the
purely kinetic models, namely the Lagrangians $\mathcal{L}=\mathcal{L}(X)$.}
It is interesting to note that, if $ \alpha \geq 0$ 
the term $\exp{\left(- \int^N dN'  \alpha(N')\right)}$ determines 
$\rho_{0}$.
In order to have $\rho > 0$ we have to require $\bar{K} > 0$. 

First of all it is worth to make 
some comments on $w$ and $c_s^2$. If
we impose the conditions $w+1 \geq 0$ and $c_s^2 \geq 0$,
in terms of $\alpha$ and $w$, or, equivalently,
of $\alpha$ and $\lambda$, the effective speed of sound, Eq.~(\ref{cs}), reads
\footnote{In purely kinetic models ($\lambda=0$), we get $(1/\alpha) d \ln X/d N \leq 0$. 
Therefore if $\alpha > 0$, $X$ can only decrease with time 
to its minimum value \cite{Bertacca:2007ux}.}   
\begin{equation}
c_s^2 =-\frac{(w+1)}{2\alpha} \frac{d \ln X}{d N} 
= -\frac{\alpha - \lambda}{6\alpha} \frac{d \ln X}{d N}\ge 0 \; .
\end{equation}

If the universe is dominated by a
fluid with equation of state $w_B={\rm const}$ then 
\begin{equation}
H=\dot{N} \sim 2/[3(w_B+1)t]
\end{equation}
and, if  $\alpha \neq 0$ and $f \sim \varphi^{-\beta}$,
up to a multiplicative constant, we have
\begin{equation}
\label{alpha_Wb_constant}
\alpha(t) = 3(w+1) - \frac{3}{2}\beta (w_{B}+1)\frac{\sqrt{2X} t}{\varphi}\;.
\end{equation}
When $\alpha=0$, we recover the scaling k-essence models 
\cite{Bertacca:2007ux,chiba:2000}.

Now we want to describe some properties of the Lagrangian studied in 
{\it Example 2)} of Section \ref{BIL0} (see also Ref.~\cite{Bertacca:2007ux})
when the universe is dominated by the radiation ($w_B=w_R=1/3$). 
Specifically, we want to investigate how 
our UDM fluid behaves during this epoch 
and for what values of $\varphi$ the kinetic term $X$ provides
a basin of attraction.
Moreover, for the sake of simplicity we impose $M^4=\Lambda$ and 
we apply the field redefinition  
$\sqrt{3}\varphi/2\rightarrow\sqrt{3}\varphi/2-\pi/2$.
Then, for $\varphi>0$, the Lagrangian becomes
\begin{equation}
\label{udm-sin}
{\mathcal L} = - \Lambda \frac{\sqrt{1-2X/\Lambda} }
{\sin \left(\frac{\sqrt{3}}{2} \varphi \right)} \;. 
\end{equation}
At early times, 
$\sqrt{3}\varphi(a \sim 0)/2 \ll 1$
and $2X(a \sim 0)/\Lambda \approx 1$. Therefore we have $\beta=1$ and 
$ w\sim 0$. 
Therefore from Eq.~(\ref{alpha_Wb_constant}) we get 
\begin{equation}
\alpha(t) =3-2\frac{\sqrt{2X} t}{\varphi}\;.
\end{equation}
By recalling the definition of $\alpha$, 
\begin{equation}
\label{alpha}
\alpha=-\frac{d\ln[2X(\partial g/\partial X)-g]}{d N}\;,
\end{equation}
from the Born-Infeld type kinetic term, we obtain 
\begin{equation}
\label{EqX_Wb_constant}
-\frac{2\dot{X}/\Lambda}{1-2X\Lambda}t=3-2\frac{\sqrt{2X} t}{\varphi}\;.
\end{equation}
Now it is easy to see that if $\varphi\simeq \sqrt{2X}t$, at early times 
the variation of the kinetic term is slow, as required to obtain appropriate
values of $X$ and $\varphi$ when the universe enters the UDM-dominated 
epoch. 
In fact, by solving the differential equation (\ref{EqX_Wb_constant}), 
we obtain
\begin{equation}
\label{X-rad}
X=\frac{\Lambda}{2}(1-\xi t)\;,
\end{equation}
where $\xi$ is a positive integration constant. By hypothesis, we know 
that $2X/\Lambda \approx 1$
then $\xi \ll 1$ and $\varphi\simeq \sqrt{\Lambda}t$. Therefore $\varphi$ and 
$X$ vary slowly and the solution is sufficiently stable during the 
radiation-dominated epoch. We can thus determine the value of $\xi$ 
(i.e. of $X(a \sim 0)$) at early times. 

Now we want to study some properties of the initial conditions of our
UDM fluid.
First of all we want to know for what values of $\varphi$ 
we can have a basin of attraction in $X$.
By making explicit $w$ in terms of $g(X)$, we rewrite the relation 
(\ref{sigma}) as
\begin{equation}
\label{alphaX}
\alpha(t) = -\frac{2\dot{X}/\Lambda}{1-2X\Lambda}t=6X/\Lambda
-2\frac{\sqrt{2X} t}{\varphi}\;.
\end{equation}
If $\alpha<0$ then $\dot{X}>0$. Thus, with the help of Eq.~(\ref{alphaX}), we
get $2X_{\rm in}/\Lambda<[2\sqrt{\Lambda}t_{\rm in}/(3\varphi_{\rm in})]^2$ 
and therefore, for a given $t_{\rm in}$, we can impose a suitable value 
of $\varphi_{\rm in}$ such that this condition is satisfied.
Specifically, we impose that $3\varphi_{\rm in}<2\sqrt{\Lambda}t_{\rm in}$ and
rewrite Eq.~(\ref{alphaX}) as follows 
\begin{equation}
\dot{\chi}=\frac{(1-\chi)}{t}\left(2\sqrt{\chi}
\frac{\sqrt{\Lambda}t}{\varphi}-3\chi\right)
\end{equation}
where $\chi=2X/\Lambda$. In this case $\dot{\chi}\gg1$ because 
we are at early times and
$\chi$ starts growing very fast. We reach the condition 
$\dot{\chi}\rightarrow 0$ at some later time $\hat{t}>t_{\rm in}$.
It is important to choose $\varphi_{\rm in}$ 
such that $\hat{\chi}=2X(\hat{t})/\Lambda \simeq 1$ and 
$2\sqrt{\Lambda}\hat{t}/\hat{\varphi}-3\ge0$, where 
$\hat{\varphi}=\varphi(\hat{t})=\sqrt{\Lambda}\hat{t}-K$ 
and $\sqrt{\Lambda}\hat{t}\le 3 K < 3\sqrt{\Lambda}\hat{t}$. Obviously, 
$K$ depends on $\varphi_{\rm in}$.
Finally, for $t\ge\hat{t}$, $\alpha\to 0$ and then becomes positive. 
Consequently, $\dot{X}<0$ and we recover the solution studied previously, 
i.e. Eq.~(\ref{X-rad}). 
 
\subsection{Lagrangian of the type $\mathcal{L}(\varphi ,X)=
f(\varphi)g(X)-V(\varphi)$}
\label{rad_fg-V}
Starting from the scalar field Lagrangian considered 
in Eq. (\ref{Lagrangian}),
the energy density $\rho$, the equation of state $w$ and 
the speed of sound $c_s^2$ are given by Eqs. (\ref{rho_fg-v}), 
(\ref{w_fg-v}) and (\ref{c_s^2_fg-v}), respectively.
We can write the equation of motion (\ref{eq_phi_fg-v}) as follows
\begin{eqnarray}
 \label{eq_phi_fg-v2}
\fl
\frac{1}{f}\frac{df}{dN}= \lambda_1(N,X) \nonumber \\
\fl
\frac{1}{f}\frac{dV}{dN}= \lambda_2(N,X) \left(2X\frac{\partial g}{\partial X}
- g\right)\nonumber \\
\fl
\left(\frac{\partial g}{\partial X} +  2X\frac{\partial^2 g}{\partial X^2}
\right) dX
+ \left[ 3 \left(2 X \frac{\partial g}{\partial X}\right) +
\lambda(N,X) \left(2 X \frac{\partial g}{\partial X} - g \right) 
\right] dN = 0 \;, 
\end{eqnarray}
where also in this case $\lambda=\lambda_1+\lambda_2$ is  a generic 
function of $N$ and $X$ and $\lambda_2=(dV/dN)/(\rho-V)$. 
Now, following the same reasoning of \ref{fg}, we obtain
\begin{equation}
\label{sigma_fg-v}
\alpha - \lambda = 3 (w_V+1)=3\frac{2X(\partial g/\partial X)}{2X(\partial 
g/\partial X) - g}\;,
\end{equation}
where $\alpha$ is given by Eq. (\ref{alpha}) and $w_V=(p+V)/(\rho-V)$, and 
we recover Eq.~(\ref{general_solution}). Then
\begin{equation}
\label{V}
V(N)=V_0+\bar{K}\int_{N_0}^N \left[\lambda_2(N')e^{ - \int^{N'} dN'' 
\alpha(N'')}dN'\right]
\end{equation}
and,
\begin{equation}
\label{density_k}
\fl
 \rho =V_0+\bar{K}\left\{e^{ - \int^N dN' \left( \alpha(N')-
\lambda_1(N')\right)}+\int_{N_0}^N \left[\lambda_2(N')e^{ - \int^{N'} dN'' 
\alpha(N'')}dN'\right]\right\}\;.
\end{equation}
In other words, the quantities $\alpha$, $\lambda_1$ and $\lambda_2$ 
completely describe the dynamics of these models.

Now, in the radiation-dominated epoch, $\lambda_2(a\sim0)\propto a^3~ 
dV/dN(a\sim0) \simeq0$.
Therefore, by considering the Lagrangian of the model 
in Section \ref{BIL}, by defining $M^4=\Lambda$ and knowing that 
$\sqrt{3}\varphi(a \sim 0)/2 \ll (1-c_\infty^2)$
and $2X(a \sim 0)/\Lambda \sim 1$, we immediately recover the
particular case investigated in \ref{fg}.

%%%%%%%%%%%%%%%%%%%%%%%%%%%%%%%%%%%%%%%%%%%%%%%%%%%%%%%%%%%%%%%%%%%%%%%%%%%%%
\def\theequation{C.\arabic{equation}}
\vskip 0.2cm
\section{Proof of the equivalence of Lagrangians 
of type $\mathcal{L}=\mathcal{L}(h(\varphi)X,\varphi)$}
\label{GeneralClass-eq}

We briefly investigate some properties of invariance
of the Lagrangians $\mathcal{L}=\mathcal{L}(h(\varphi)X,\varphi)$.
Write the equation of motion as follows
 \begin{equation}
 \label{eq_phi2}
\left(\frac{\partial \mathcal{L}}{\partial X} 
+2X\frac{\partial^2 \mathcal{L}}{\partial X^2}\right) \dot X
+\dot \varphi \frac{\partial}{\partial \varphi}\left(2X\frac{\partial 
\mathcal{L}}{\partial X}
- \mathcal{L}\right)=-6HX\frac{\partial \mathcal{L}}{\partial X} \;, 
\end{equation}
where the RHS of Eq.~(\ref{eq_phi2}) is an explicit function of time, 
through $H$, and has the meaning of ``non-inertial force'' in the 
equation of motion.

In particular we want to prove that we can always make the following 
change of field variable
\begin{equation}
\label{hX=RZ}
h(\varphi)X=\mathcal{R}(\theta)Z\;,\quad\quad {\rm where}\quad\quad Z=
\dot{\theta}^2/2\;.
\end{equation}
Consider $h(\varphi)>0$ and $\mathcal{R}(\theta)>0$ are continuous functions 
of $\varphi$ and $\theta$ respectively, with $h(\varphi)\neq 
\mathcal{R}(\theta)$.
Eq.~(\ref{hX=RZ}) can be written in the following differential form
\begin{equation}
h(\varphi)^{1/2}d\varphi\mp\mathcal{R}(\theta)^{1/2}d\theta=0\;.
\end{equation}
Without any loss of generality, hereafter we consider the case with the minus 
sign.

We have indirectly constructed a map 
$\iota:\varphi \rightarrow \iota(\varphi)$ such that
\begin{equation}
\theta\equiv\iota(\varphi)=\int_{\varphi_0}^{\varphi}
\left\{\left[\frac{h(\tilde{\varphi})}
{\mathcal{R}(\tilde{\varphi})}\right]^{1/2}d\tilde{\varphi}\right\}
+\theta_0\;,
\end{equation}
where $\mathcal{R}(\theta)=\mathcal{R}(\iota(\varphi))\equiv 
\mathcal{R}(\varphi)$. 
We necessarily have 
$h(\varphi)=h(\iota^{-1}(\theta))\equiv h(\theta)$.
Therefore, the Lagrangian becomes 
$\mathcal{L}(h(\varphi)X,\varphi)=
\mathcal{L}(\mathcal{R}(\theta)Z,\iota^{-1}(\theta))=
\mathcal{L}(\mathcal{R}(\theta)Z,\theta)$.
In order to study the change of field variables, we rewrite 
Eq.~(\ref{hX=RZ}) in differential form
\begin{equation}
X\left(\frac{\partial h}{\partial \varphi}\right)d\varphi+hdX=
Z\left(\frac{\partial \mathcal{R}}{\partial \theta}\right)d\theta+
\mathcal{R}dZ\;.
\end{equation}
Then
\begin{equation}
\label{dotX}
\dot X=Z\left[\frac{\partial}{\partial \theta}\left( 
\frac{\mathcal{R}}{h}\right)\right]\dot\theta
+\left(\frac{\mathcal{R}}{h}\right)\dot Z\;.
\end{equation}
Finally, starting from this change of field variables, 
we are able to prove that their equations of motion are dynamically 
equivalent, namely 
Eq.~(\ref{eq_phi2}) is identical to
\begin{equation}
 \label{eq_theta}
\left(\frac{\partial \mathcal{L}}{\partial Z} 
+2Z\frac{\partial^2 \mathcal{L}}{\partial Z^2}\right) \dot Z
+\dot \theta \frac{\partial}{\partial \theta}
\left(2Z\frac{\partial \mathcal{L}}{\partial Z}
- \mathcal{L}\right)=-6HZ\frac{\partial \mathcal{L}}{\partial Z} \;, 
\end{equation}
and that they consequently have the same 
equation of state and effective speed of sound, i.e. Eqs.~(\ref{w})
and (\ref{cs}) are respectively equal to
\begin{equation}
\label{wcs}
w = \frac{\mathcal{L}}{2Z \frac{\partial \mathcal{L}}{\partial Z} - 
\mathcal{L}} \;, \quad \quad {\rm and} \quad\quad c_s^2 = \frac{(\partial 
\mathcal{L} /\partial Z)}
{(\partial \rho/\partial Z)} = 
\frac{\frac{\partial \mathcal{L} }{\partial Z}}{\frac{\partial Z}
{\partial Z}+ 2Z\frac{\partial^2 \mathcal{L}}{\partial Z^2}} \;.
 \end{equation}

The proof is a trivial consequence of Eqs.~(\ref{hX=RZ}),(\ref{dotX}) 
and the following relations
\begin{equation}
 \frac{\partial \mathcal{L}}{\partial X}=  
\left(\frac{h}{\mathcal{R}}\right)\frac{\partial \mathcal{L}}{\partial Z}\;, 
\quad \quad
\frac{\partial^2 \mathcal{L}}{\partial X^2}=
\left(\frac{h}{\mathcal{R}}\right)^2\frac{\partial^2 
\mathcal{L}}{\partial Z^2}\;,
\end{equation}
\begin{equation}
\frac{\partial \mathcal{L}}{\partial \varphi}= 
\left(\frac{h}{\mathcal{R}}\right)^{1/2}
\frac{\partial \mathcal{L}}{\partial \theta}+Z\left( 
\frac{\mathcal{R}}{h}\right)^{1/2}
\left[\frac{\partial}{\partial \theta}\left( 
\frac{h}{\mathcal{R}}\right)\right]
\frac{\partial \mathcal{L}}{\partial Z}\;, 
\end{equation}
\begin{equation}
\frac{\partial^2 \mathcal{L}}{\partial X\partial \varphi}=
 \left(\frac{h}{\mathcal{R}}\right)^{1/2}\left\{\left[\frac{\partial}{\partial
 \theta}
\left( \frac{h}{\mathcal{R}}\right)\right]
\left(\frac{\partial \mathcal{L}}{\partial Z} +Z\frac{\partial^2 \mathcal{L}}
{\partial Z^2}\right)+\left( \frac{h}{\mathcal{R}}\right)\frac{\partial^2 
\mathcal{L}}
{\partial Z\partial \theta}\right\}\;.
\end{equation}
If $\mathcal{R}(\theta)=1$ (or $h(\varphi)=1$) we can immediately recover the
particular case investigated in Section \ref{Class-eq}.

%%%%%%%%%%%%%%%%%%%%%%%%%%%%%%%%%%%%%%%%%%%%%%%%%%%%%%%%%%%%%%%%%%%%%%%%
\def\theequation{D.\arabic{equation}}
\vskip 0.2cm
\section{Other examples of coordinate transformations of
Lagrangians with a Born-Infeld kinetic term}
\label{CoordTransform}

Here, we give some examples of how a coordinate transformation 
of Lagrangians with a Born-Infeld kinetic term can yield
Lagrangians with different kinematical properties. 
In fact, starting from the equality (\ref{LXphi-Ztheta}) 
and still with a Born-Infeld kinetic term,
we can see that if $Z=f(\varphi) X\;$, and 
$\mathcal{W}(\theta)=V(\varphi)+f(\varphi)$, we obtain
\begin{equation}
\mathcal{L}=-f(\theta)\left[1-\frac{2Z/M^4}{f(\theta)}\right]^{1/2}
+f(\theta)-\mathcal{W}(\theta)\;,
\end{equation}
where we have assumed that $\mathcal{W}(\theta)>0$. 
In other words, it is possible to transform a Born-Infeld Lagrangian 
into a Dirac-Born-Infeld Lagrangian \cite{Alishahiha:2004eh}.
This is a particular case of a more general transformation.
In fact, if $X=Z/\mathcal{T}(\theta)$ and $V(\varphi)=\mathcal{W}(\theta)
-\mathcal{T}(\theta)$~(with $\mathcal{W}(\theta)>0$
and $\mathcal{T}(\theta)>0$),
we get 
\begin{equation}
\label{Lfg(Z/T)+T-W}
\mathcal{L}=-f(\theta)\left[1-\frac{2Z/M^4}{\mathcal{T}(\theta)}\right]^{1/2}+
\mathcal{T}(\theta)-\mathcal{W}(\theta)\;. 
\end{equation}

Starting from the model of Section \ref{BIL}~, we can obtain, for 
example, two similar Lagrangians that can be rewritten in the form 
(\ref{Lfg(Z/T)+T-W}). Define 
\begin{equation}
\mathcal{T}^{1/2}[\theta(\varphi)]=\kappa_i^{1/2} 
\frac{\cosh{(\gamma \varphi)}}{[1+(1- c_\infty^2)\sinh^2{(\gamma 
\varphi)}]^{1/2}}\;,
\end{equation}
where $i=1,2$ and $\kappa_1=\Lambda c_\infty^2/(1- c_\infty^2)$ and 
$\kappa_2=\Lambda/(1- c_\infty^2)$.
In this case, 
\begin{equation}
\theta(\varphi)=\frac{1}{\gamma}
\left(\frac{\kappa_i}{1- c_\infty^2}\right)^{1/2}
{\rm arc}\sinh\left[(1-c_\infty^2)^{1/2}\sinh{(\gamma \varphi)}\right]\;,
\end{equation}
and the various terms of Eq.~(\ref{Lfg(Z/T)+T-W}) become
\begin{equation}
f(\theta)=\frac{\Lambda c_\infty}{1- c_\infty^2}
\frac{\left\{1+(1- c_\infty^2)\sinh^{-2}{\left[
\left(\frac{1- c_\infty^2}{\kappa_i}\right)^{1/2}\gamma 
\theta\right]}\right\}^{1/2}}
{\cosh^2{\left[\left(\frac{1- c_\infty^2}{\kappa_i}\right)^{1/2}\gamma 
\theta\right]}}\;,
\end{equation}
\begin{equation}
\mathcal{T}(\theta)=\frac{\kappa_i}{(1- c_\infty^2)}
 \frac{1- c_\infty^2 + 
\sinh^2{\left[\left(\frac{1- c_\infty^2}{\kappa_i}\right)^{1/2}\gamma 
\theta\right]}}
{\cosh^2{\left[\left(\frac{1- c_\infty^2}{\kappa_i}\right)^{1/2}\gamma 
\theta\right]}}\;,
\end{equation}
and
\begin{equation}
\fl
\mathcal{W}(\theta)=\frac{\Lambda}{1- c_\infty^2}
\frac{({1- c_\infty^2)\cosh^2{\left[\left(\frac{1- 
c_\infty^2}{\kappa_i}\right)^{1/2}\gamma \theta\right]}}
+(\kappa_i/\Lambda)\sinh^2{\left[\left(\frac{1- 
c_\infty^2}{\kappa_i}\right)^{1/2}\gamma \theta\right]}}
{\cosh^2{\left[\left(\frac{1- c_\infty^2}{\kappa_i}\right)^{1/2}\gamma 
\theta\right]}}\;.
\end{equation}

%%%%%%%%%%%%%%%%%%%%%%%%%%%%%%%%%%%%%%%%%%%%%%%%%%%%%%%%%%%%%%%%%%%%%%%%
%%%%%%%%%%%%%%%%%%%%%%%%%%%%%%%%%%%%%%%%%%%%%%%%%%%%%%%%%%%%%%%%%%%%%%

\end{document}